\documentclass[showpacs,twocolumn,floatfix]{revtex4}
\usepackage{graphicx}
\usepackage{amsmath}

\input{tcilatex}

\begin{document}

\title{Proposal for teleportation of charge qubits via superradiance}
\author{Y. N. Chen$^{1}$, C. M. Li$^{1}$, D. S. Chuu$^{1}$, and T. Brandes$%
^{2}$}
\affiliation{$^{1}$Department of Electrophysics, National Chiao-Tung University, Hsinchu
300, Taiwan\\
$^{2}$School of Physics and Astronomy, The University of Manchester P.O. Box
88, Manchester, M60 1QD, U.K.}
\date{\today}

\begin{abstract}
A scheme is proposed to teleport charge qubits via superradiance.
Reservoir-induced entanglement is generated between two semiconductor dots
in a microcavity where a quantum state encoded in a third quantum dot is
then tuned into collective decay with one of the entangled dots.
Teleportation is achieved automatically in our scheme which we also extend
to quantum wires.
\end{abstract}

\maketitle

\bigskip

Quantum entanglement has achieved a prime position in current research due
to its central role in quantum information science, e.g., in quantum
cryptography, quantum computing, and teleportation \cite{1}. Many efforts
have been devoted to the study of entanglement induced by a direct
interaction between the individual subsystems. Recently, attention has been
focused on \emph{reservoir-induced} entanglement \cite{2} with the purpose
of shedding light on the generation of entangled qubits at remote separation.

Entangled states can also be generated via sub- and superradiance, i.e. the
collective spontaneous decay first introduced by Dicke \cite{3}. For the
simplest case of two identical two level atoms interacting with the vacuum
fluctuations of a common photon reservoir, entanglement naturally appears in
the two intermediate states 
\begin{equation}
\left| T_{0}\right\rangle =\frac{1}{\sqrt{2}}(\left| \uparrow \downarrow
\right\rangle +\left| \downarrow \uparrow \right\rangle ),\quad \left|
S_{0}\right\rangle =\frac{1}{\sqrt{2}}(\left| \uparrow \downarrow
\right\rangle -\left| \downarrow \uparrow \right\rangle )
\end{equation}%
of the two decay channels from the excited state $\left| T_{1}\right\rangle
=\left| \uparrow \uparrow \right\rangle $ to the ground state $\left|
T_{-1}\right\rangle =\left| \downarrow \downarrow \right\rangle $. An
experimental demonstration of two-ion collective decay as a function of
inter-ion separation was shown by Devoe and Brewer in 1996 \cite{4}. On the
other hand, the possibility to modify decay rates of individual atoms inside
cavities (Purcell effect) \cite{5} has been known for a long time, and
enhanced and inhibited spontaneous rates for atomic systems were intensively
investigated in the 1980s \cite{6} by passing atoms through cavities.

Experiments of teleportation have already been realized in NMR \cite{7},
photonic \cite{8}, and atomic \cite{9} systems. Turning to solid state
systems, however, experimental demonstration of teleportation in charge
qubits is still lacking, and only few theoretical schemes are proposed. \cite%
{10} In this work, we propose a teleportation scheme for atomic and solid
state qubits (quantum dots (QD) and quantum wires), which is based on the
Dicke effect and achieves, in contrast to usual schemes, a
''one-pass''teleportation by a joint measurement.

\begin{figure}[t]
\includegraphics[width=7cm]{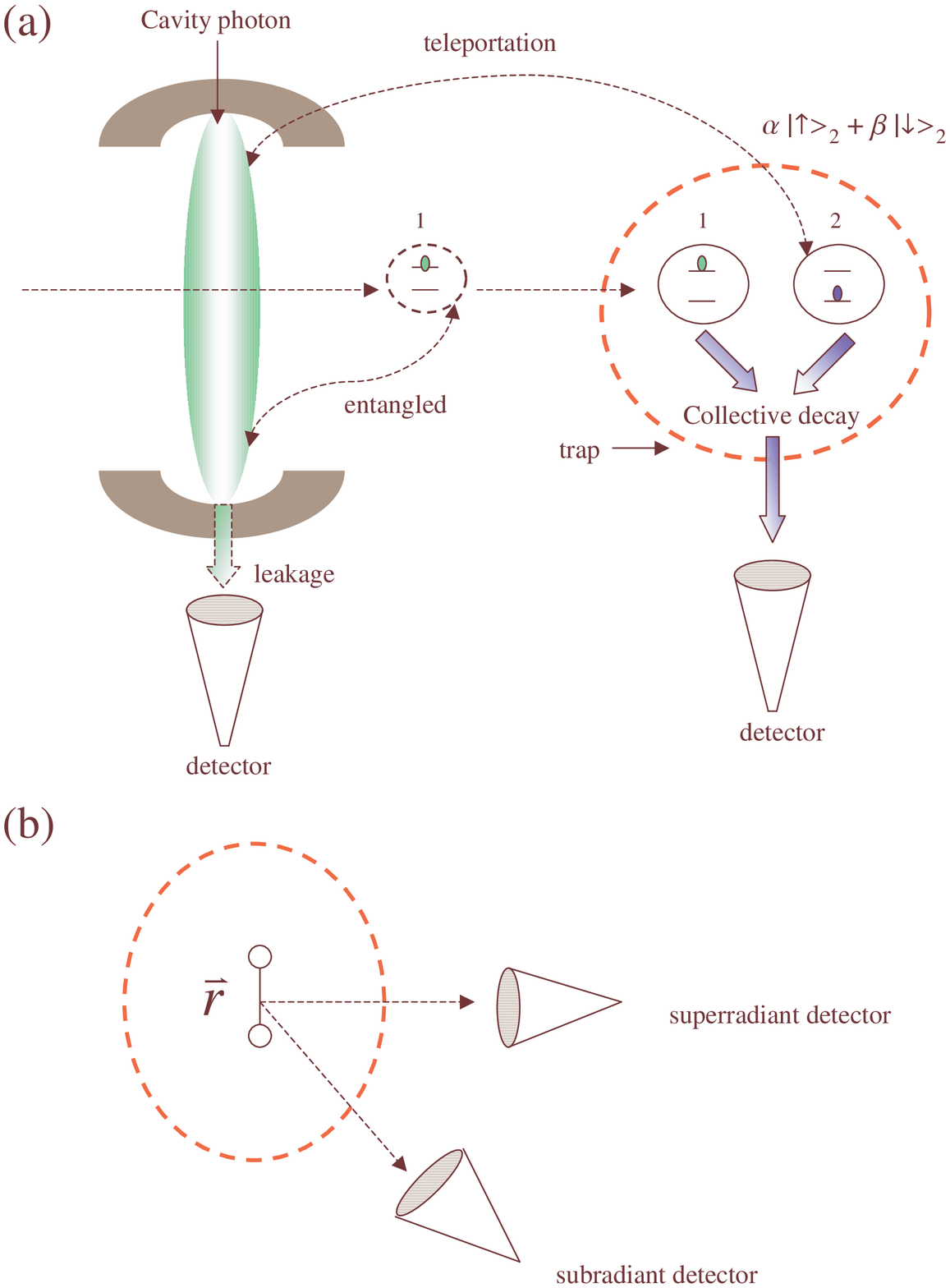}
\caption{{}(a) Schematic description of teleportation by collective decay in
a cavity QED system. First, the singlet entangled state is generated between
atom-1 and the cavity photon as the atom has passed through the cavity. Atom
1 and 2 then decay collectively. If the measurement outcome is a single
photon with subradiant decay rate, the teleportation to the cavity photon
state is achieved automatically. (b) To distinguish between super- and
subradiant decay, two detectors are placed at appropriate angles such that
the emitted photon momentum $\protect\overrightarrow{k}$ satisfy the
condition of $\protect\overrightarrow{k}\cdot $ $\protect\overrightarrow{r}=0
$ and $\protect\pi $, respectively.}
\end{figure}

\emph{The simplest case}. --- To address the role of superradiance in
teleportation, let us first consider a two-level atom passing through a
cavity as shown in Fig.1. In the strong coupling regime, the interaction
between the atom and the cavity photon is%
\begin{equation}
H^{\prime }=\hbar g(\sigma _{+}b^{-}+\sigma _{-}b^{+}),
\end{equation}%
where $g$ is the atom-cavity coupling strength, $b^{\pm }$ and the Pauli
matrices $\sigma ^{\pm }$ are the cavity photon and atom operators,
respectively. With the appropriate preparation of the initial state of
atom-1 and the control of its passing time through the cavity, the singlet
entangled state $\frac{1}{\sqrt{2}}(\left| 0\right\rangle _{c}\left|
\uparrow \right\rangle _{1}-\left| 1\right\rangle _{c}\left| \downarrow
\right\rangle _{1})$ is created between atom-1 and the cavity photon. The
notations $\left| \uparrow \right\rangle _{1}$ ($\left| \downarrow
\right\rangle _{1}$) and $\left| 0\right\rangle _{c}$ ($\left|
1\right\rangle _{c}$) refer, respectively, to atom-1 in the excited (ground)
state and no (one) photon in the cavity. If the quantum state $\alpha \left|
\uparrow \right\rangle _{2}+\beta \left| \downarrow \right\rangle _{2}$ ,
generated by the coherent excitation of atom-2 with a laser pulse, is to be
teleported, the next step to complete the teleportation is to trap both
atoms close to each other. In this case, the total wave function of the
system is given by 
\begin{eqnarray}
&&\left| \Psi \right\rangle =\frac{1}{\sqrt{2}}(\left| 0\right\rangle
_{c}\left| \uparrow \right\rangle _{1}-\left| 1\right\rangle _{c}\left|
\downarrow \right\rangle _{1})\otimes (\alpha \left| \uparrow \right\rangle
_{2}+\beta \left| \downarrow \right\rangle _{2})  \notag \\
&=&\left| 0\right\rangle _{c}\otimes (\frac{\alpha }{\sqrt{2}}\left|
T_{1}\right\rangle _{12})+\left| 1\right\rangle _{c}\otimes (\frac{-\beta }{%
\sqrt{2}}\left| T_{-1}\right\rangle _{12}) \\
&+&(\alpha \left| 1\right\rangle _{c}+\beta \left| 0\right\rangle
_{c})\otimes \frac{|S_{0}\rangle _{12}}{2}+(-\alpha \left| 1\right\rangle
_{c}+\beta \left| 0\right\rangle _{c})\otimes \frac{|T_{0}\rangle _{12}}{2}.
\notag
\end{eqnarray}%
Since atom-1 and 2 are now close enough, the common photon reservoir will
drive them to decay collectively with four possibilities for the detector's
results: zero photon, two photons, one photon via the superradiant channel,
or one photon via the subradiant channel. If the measurement outcome is a
single photon with a suppressed decay rate, the teleportation is achieved
automatically. As for the result of one photon with enhanced decay rate, all
we have to do is to perform a phase-gate operation on the cavity photon
state to complete the teleportation.

Since decay time is a statistical average, one might ask how to distinguish
between sub- and superradiant photons via the decay time in one single shot
? We would like to point out that because of the collective decay, the
momentum of the emitted photon $\overrightarrow{k}$ depends on the
separation of the two atoms $\overrightarrow{r}$, i.e. $\overrightarrow{k}%
\cdot $ $\overrightarrow{r}=0$ or $\pi $ corresponds to the emission of
super- or sub-radiant photon, respectively. \cite{4} Therefore, sub- and
super-radiance can be distinguished by placing detectors at appropriate
angles as shown in Fig. 1 (b). After monitoring all possible emission
directions, one can further make sure whether the number of the emitted
photon is one or not. The teleportation can then be tested by repeating this
scheme over many cycles and probing the state of the cavity (one or no
photon) after each cycle.

\begin{figure}[h]
\includegraphics[width=7.5cm]{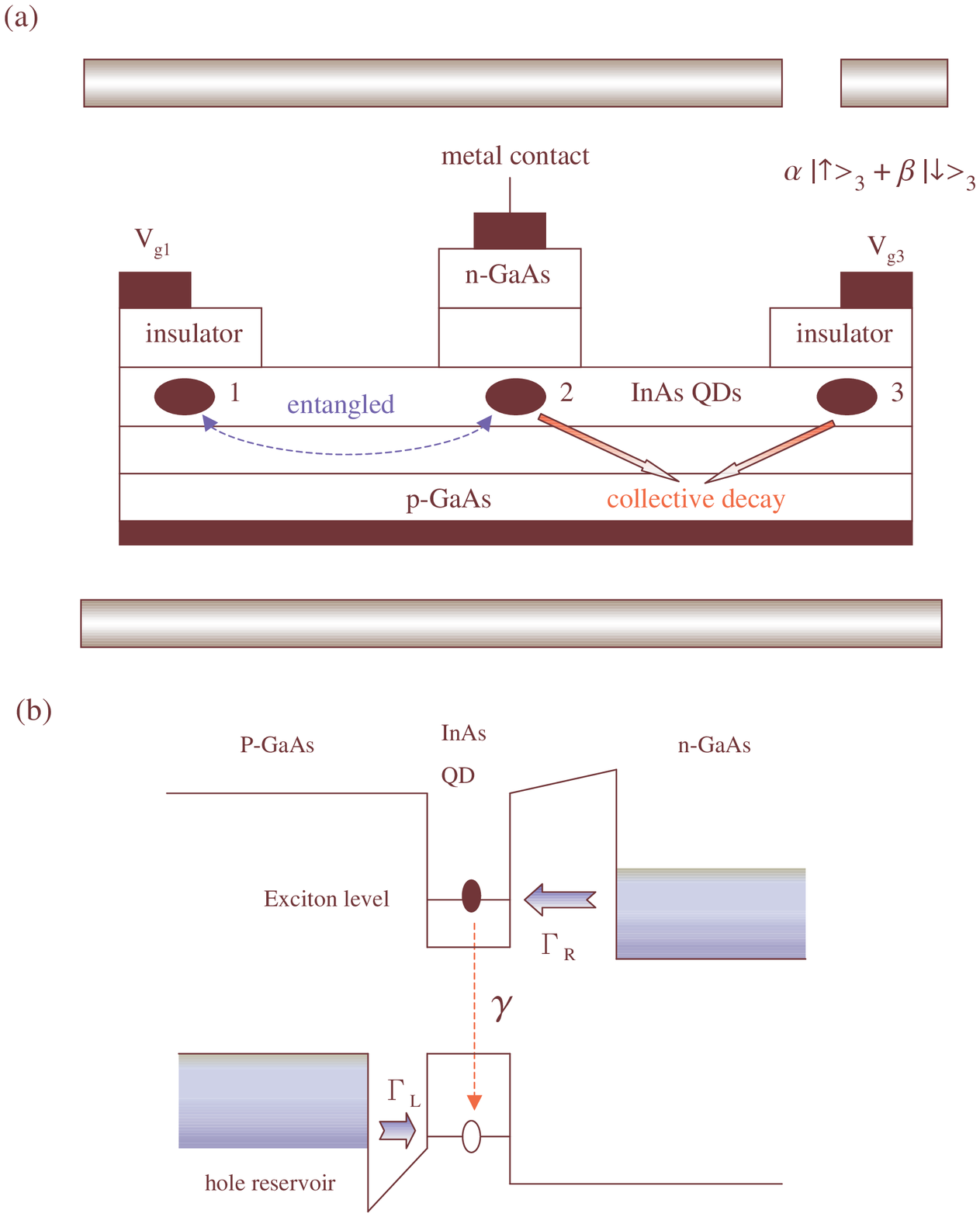}
\caption{{}(a) Schematic description of teleportation by the Dicke effect in
quantum dots. First, a subradiance-induced singlet entangled state is
generated between QD 1 and 2. The energy bandgap of the exciton in QD 3 (1)
is then tuned to be (non)-resonant with that in QD 2. Finally, a joint
measurement is done naturally by collective decay of QD 2 and 3. (b) Energy
band diagram of an InAs QD embedded inside a p-i-n junction.}
\end{figure}

\emph{Quantum dots}. --- We now proceed to consider three semiconductor
quantum dots embedded inside a p-i-n junction as shown in Fig. 2. The Fermi
level of the \textit{p(n)}-side hole (electron) is assumed to be slightly
lower (higher) than the hole (electron) subband in the dot. After a hole is
injected into the hole subband in the quantum dot, the \textit{n}-side
electron can tunnel into the exciton level because of the Coulomb
interaction between the electron and hole. The effective Hilbert space of
the closed system is spanned by $\left| \uparrow \right\rangle =\left|
e,h\right\rangle $ and $\left| \downarrow \right\rangle =\left|
0,0\right\rangle $, which correspond to the exciton and ground state in the
quantum dot. \cite{11} The interactions between the $i$-th QD and the
reservoirs can then be written as 
\begin{equation}
H_{V}=\sum_{\mathbf{q},i}(V_{\mathbf{q}}c_{\mathbf{q}}^{\dagger }\overset{%
\wedge }{s_{R,i}}+W_{\mathbf{q}}d_{\mathbf{q}}^{\dagger }\overset{\wedge }{%
s_{L,i}}+\text{H.c.}),
\end{equation}%
where $\overset{\wedge }{s_{R}}=\left| h\right\rangle \left\langle \uparrow
\right| $, $\overset{\wedge }{s_{L}}=\left| h\right\rangle \left\langle
\downarrow \right| $, $V_{\mathbf{q}}$ and $W_{\mathbf{q}}$ couple the
channels $\mathbf{q}$ of the electron and the hole reservoirs, and the extra
state $\left| h\right\rangle $ means there is one hole in the quantum dot.
Here, $c_{\mathbf{q}}$ and $d_{\mathbf{q}}$ denote the electron operators in
the left and right reservoirs, respectively. The state $\left|
e\right\rangle $ can be neglected by using a thicker barrier on the electron
side so that there is little chance for an electron to tunnel in advance.

If the energy level differences of the first and second dot excitons are
tuned to be resonant with each other, superradiant decay occurs between them
with the interaction 
\begin{eqnarray}
H_{P} &=&\sum_{\mathbf{k}}\frac{1}{\sqrt{2}}\{D_{\mathbf{k}}b_{\mathbf{k}%
}[(1+e^{i\mathbf{k}\cdot \mathbf{r}})\left| S_{0}\right\rangle \left\langle
T_{-1}\right|   \notag \\
&&+(1-e^{i\mathbf{k}\cdot \mathbf{r}})\left| T_{0}\right\rangle \left\langle
T_{-1}\right| ]+\text{H.c.}\},
\end{eqnarray}%
where $\left| S_{0}\right\rangle =\frac{1}{\sqrt{2}}(\left| \uparrow
\right\rangle _{1}\left| \downarrow \right\rangle _{2}-\left| \downarrow
\right\rangle _{1}\left| \uparrow \right\rangle _{2})$ and $\left|
T_{0}\right\rangle =\frac{1}{\sqrt{2}}(\left| \uparrow \right\rangle
_{1}\left| \downarrow \right\rangle _{2}+\left| \downarrow \right\rangle
_{1}\left| \uparrow \right\rangle _{2})$ represent the singlet and triplet
entangled states in these two QDs (1 and 2). Here, $b_{\mathbf{k}}$ is the
photon operator, $D_{\mathbf{k}}$ is the coupling strength, and $\mathbf{r}$
is the position vector between dot 1 and 2.

The transient behavior of the non-equilibrium problem caused by the
interactions $H_{V}$ and $H_{P}$ can be solved by master-equation
techniques. If the current is conducted through dot 2 only, however, maximum
entangled states can be created in the long time limit and for the system
being inside a rectangular microcavity: it turns out that for small
radiative decay rates, the steady state density operator becomes independent
of the coupling to the leads and approaches a pure singlet or triplet
entangled state depending on the inter-dot distance \cite{12}. This means
that even for remote separation of the dots and under transport conditions,
the entanglement can still be achieved.

Once the entangled state is formed, we then lower the energy bandgap of the
QD 1 exciton using the Stark effect, such that its level spacing becomes
smaller than the lowest mode of the cavity photon. Thus, the interaction
between the photon and the exciton in QD 1 is turned off. To prevent further
decoherence of the exciton, the couplings to the left and right electron
reservoirs should also be switched off. If the quantum state $\alpha \left|
\uparrow \right\rangle _{3}+\beta \left| \downarrow \right\rangle _{3}$ ,
generated by the coherent excitation of a laser pulse \cite{13}, in QD 3 is
to be teleported, the next step to complete the teleportation is to tune the
exciton in QD 3 to be resonant with that in dot 2. As we mentioned above, if
the measurement outcome is a single photon with a suppressed (enhanced)
decay rate, the teleportation is achieved automatically.

Unlike the spatially resolved scheme of superradiant/subradiant photon in
purely quantum optic systems, a time resolved measurement scheme is needed
for the QDs in a microcavity. In this case, it is possible to have the
vanishing subradiant decay rate if QD-2 and QD-3 are placed at certain
positions \cite{12}, rendering the indistinguishable from the $\left|
T_{-1}\right\rangle $ state. Therefore, in the design of the device, the
separation between QD-2 and QD-3 should not be positioned exactly equal to
the multiple (effective) wavelength of the emitted photon. \cite{14} On the
other hand, however, the time resolved measurement scheme is actually a
statistical average. To distinguish between these two channels, one has to
define a border line of time, such that above (below) it is a subradiant
(superradiant) decay. From this point of view, the ratio of superradiant to
subradiant emission rates should not be too small. This means the separation
between the QDs is better to be designed as close as to the multiple
(effective) wavelength of the emitted photon.

Based on these, we can now perform a simple estimation of the success rate:
The super- and sub-radiant decay times can be written as $(1\pm \Delta )\tau
_{0}$, where $\tau _{0}$ is the spontaneous lifetime of a single QD exciton.
The range of $\Delta $ is between 0 and 1, depending on the separation of
the QDs. Assuming that the border line is $\tau _{0}$ and the statistical
distribution of the detections on the subradiant [superradiant] emission
time is a Gaussian form centering at $(1+\Delta )\tau _{0}$ [$(1-\Delta
)\tau _{0}$] with half-width $\sigma _{-}$[$\sigma _{+}$], the probabilities
for the successful detections of $\left| T_{0}\right\rangle $ and $\left|
S_{0}\right\rangle $ states are 
\begin{eqnarray}
P_{S_{0}} &=&\frac{1}{N_{S_{0}}}\int_{\tau _{0}}^{\infty }e^{-[t-(1+\Delta
)\tau _{0}]^{2}/\sigma _{-}^{2}}dt\text{ }  \notag \\
P_{T_{0}} &=&\frac{1}{N_{T_{0}}}\int_{0}^{\tau _{0}}e^{-[t-(1-\Delta )\tau
_{0}]^{2}/\sigma _{+}^{2}}dt\text{ ,}
\end{eqnarray}%
where $N_{S_{0}}$ and $N_{T_{0}}$ are the corresponding normalization
constants. Together with Eq. (3), one can define the success rate as $%
P_{1D}\equiv (P_{S_{0}}+P_{T_{0}})/4$. For reasonable parameters of $\Delta
=0.7$ and assuming that the half-widths are equal to one-half the centering
times ($\sigma _{\pm }=(1\pm \Delta )\tau _{0}/2$), one finds that the
success rate $P_{1D}$ is about 0.47. This is a little bit lower than the
ideal success probability of 0.5. In addition to the statistical part, one
can further consider the detection efficiency $\eta $ in a real experiment
as pointed out in Ref. \cite{15}. In the detection stage, one will be able
to detect a fraction $\eta $\ of all the successful protocols. This means
one will erroneously regard a fraction $2\eta (1-\eta )$ of the cases with
two decays as successful cases. Then the success rate changes to $%
P_{suc}(\eta )=\eta P_{1D}+2\eta (1-\eta )(1-P_{1D}-P_{ND})$, where $P_{ND}$
is the probability of no decay during the detection. 

Our model can be applied to electron transport through gate-controlled
double QDs \cite{16} as well, if the quantum states are replaced as: $\left|
\uparrow \right\rangle \rightarrow \left| L\right\rangle =\left|
N_{L}+1,N_{R}\right\rangle $ and $\left| \downarrow \right\rangle
\rightarrow \left| R\right\rangle =\left| N_{L},N_{R}+1\right\rangle $,
which correspond to different many-body ground states with $N_{L}$ electrons
in the left and $N_{R}$ electrons in the right dot. For such a system, the
interaction with phonons ($\sum_{\mathbf{q}}\hbar \omega _{\mathbf{q}}a_{%
\mathbf{q}}^{\dagger }a_{\mathbf{q}}$) plays the dominant role and is given
by $H_{ph}=T_{c}[\left| L\right\rangle \left\langle R\right| X+\left|
R\right\rangle \left\langle L\right| X^{\dagger }],$ where $T_{c}$ is the
coupling strength between two dots, and $X=\prod_{\mathbf{q}}D_{\mathbf{q}}(%
\frac{\alpha _{\mathbf{q}}-\beta _{\mathbf{q}}}{\omega _{\mathbf{q}}})$ is
the phonon-dependent operator with $D_{\mathbf{q}}(z)=e^{za_{\mathbf{q}%
}^{\dagger }-z^{\ast }a_{\mathbf{q}}}$. With the advances of
nano-fabrication technologies, it is now possible to embed quantum dots in a
free standing nanosize semiconductor slab (phonon cavity). \cite{17} Just
like photon cavities, suppressed and enhanced spontaneous emission of
phonons are both possible by adjusting energy level difference of the double
dot. \cite{18} The entangled state between two qubits can also be generated
via sharing the common phonon reservoir. \cite{19}

\emph{Extension to quantum wires}. --- The present scheme can also be
applied to other solid state systems that exhibit superradiance. High
quality quantum wires in a microcavity can now be fabricated with advanced
technologies \cite{20}, with the exciton and photon forming a new eigenstate
called ''polariton'' because of the crystal symmetry in the chain direction.
The decay rate of quantum wire polaritons in a planar microcavity is \cite%
{21} 
\begin{equation}
\gamma _{1d}=\frac{2\pi e^{2}\hbar }{m^{2}c^{2}d}\sum_{n=1}^{N_{c}}\frac{1}{%
L_{c}}\frac{\left| \epsilon \cdot \chi \right| ^{2}}{\sqrt{(\frac{2\pi }{%
\lambda })^{2}-(\frac{\pi }{L_{c}}n)^{2}}},  \label{gamma1d}
\end{equation}%
where $N_{c}$ is the total number of decay channels, $\epsilon $ is the
polarization vector of the photon, $\chi $ the dipole moment of the exciton, 
$d$ the lattice spacing, and $L_{c}$ is the cavity length. Because of energy
conservation, the value of $N_{c}$ must be smaller than $2L_{c}/\lambda $,
where $\lambda $ is the wavelength of the emitted photon. If the cavity
length is designed to be roughly equal to $\lambda /2$, the Stark effect can
be applied to vary the energy gap of the exciton, and the polariton decay
can be controlled according to \ Eq.~(\ref{gamma1d}).

As for teleportation, the additional advantage over the QD systems mentioned
above is the fact that quantum wire polaritons are collective excitations
and the time-resolved measurement of many outcoming photons therefore can be
done much easier. The difference is that the preparation and readout of the
quantum states can only be accomplished by optical ways. The drawback here
is the enhancement of the renormalized frequency shift $\Omega $
(counterpart of the decay rate) by a factor of $\lambda /d$. This has a
strong effect on the entanglement and can not be neglected: assuming a
singlet entangled state is formed between the wires at $t=0$, the quantum
state of the system at $t>0$ can be written as $\Phi (t)=b_{1}(t)\left|
ex;0\right\rangle +b_{2}(t)\left| 0;ex\right\rangle +b_{0}(t)\left|
0;0\right\rangle $ with initial conditions $b_{1}(0)=\frac{1}{\sqrt{2}}$, $%
b_{2}(0)=-\frac{1}{\sqrt{2}}$, and $b_{0}(0)=0$, where ''$ex$'' and ''$0$''
represent whether or not there is an excitation in the wire. Under the
condition of short wire separations ($\ll \lambda $), the singlet state is
stable if the renormalized frequency shift is \textit{not} taken into
account. However, if it is included, the coefficients $b_{1}(t)$ and $%
b_{2}(t)$ are found to be time-dependent from the multiple-time-scale
perturbation theory \cite{22}: 
\begin{equation}
\QATOPD\{ . {\frac{d}{dt}b_{1}(t)=f_{11}b_{1}(t)+f_{12}b_{2}(t)}{\frac{d}{dt}%
b_{2}(t)=f_{21}b_{1}(t)+f_{22}b_{2}(t)},
\end{equation}%
where $f_{ij}=i\Omega _{ij}+\gamma _{ij}$, $\gamma _{ij}=\gamma _{1d}e^{i(%
\sqrt{k_{0}^{2}-k_{c}^{2}})(x_{i}-x_{j})}$, $i,j=1,2$, and 
\begin{eqnarray}
\Omega _{ij} &=&\frac{e^{2}\hbar }{m^{2}c^{2}d}\sum_{n_{c}=1}^{N_{c}}\frac{1%
}{L_{c}}\times  \\
&&\int \frac{\left| \epsilon \cdot \chi \right|
^{2}e^{ik_{x}(x_{i}-x_{j})}dk_{x}}{(k_{0}-\sqrt{k_{x}{}^{2}+k_{c}^{2}})\sqrt{%
k_{x}{}^{2}+k_{c}^{2}}}.  \notag
\end{eqnarray}%
Here, $k_{0}=2\pi /\lambda $, $k_{c}=\frac{\pi }{L_{c}}N_{c}$, and $x_{i}$
is the position of the $i$-th quantum wire. $\Omega _{ij}$ is the
position-dependent frequency shift, which contains both infrared and
ultraviolet divergence and has to be renormalized properly. \cite{23} Fig. 3
shows how the degree of entanglement (the von Neumann entropy, $E$)
decreases as a function of inter-wire separation $x$ and time $t$ when the
two wires are embedded inside a planar microcavity with cavity length $%
L_{c}=3\lambda /4$. As seen, the larger the wire separation, the faster the
decreasing of the entropy. This tells us, to maintain high fidelity of
teleportation, the quantum wires have to be fabricated as close as possible,
and the collective measurement has to be performed immediately once the
singlet state is created. Otherwise, the enhanced frequency shift will
inevitably diminish the fidelity of teleportation. 
\begin{figure}[h]
\includegraphics[width=7.5cm]{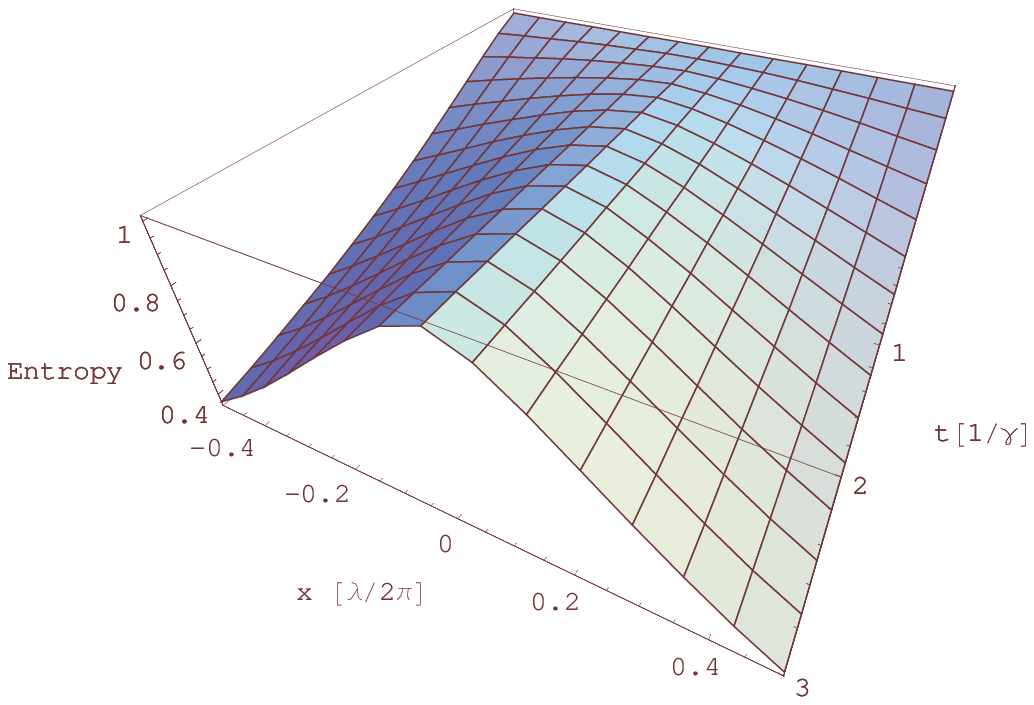}
\caption{Effect of renormalized frequency shift on the degree of
entanglement (the von Neumann entropy). }
\end{figure}

A few remarks about the comparisons between our scheme and other proposals
should be mentioned here. In usual teleportation scheme, one has to perform
Hadamard and CNOT transformations on one of the entangled particles and the
teleported quantum state. After that, the information from the joint
measurements of the two particles has to be sent to the other entangled
particle in order to allow proper unitary operations. In our proposal,
however, the Hadamard and CNOT transformations are omitted and the joint
measurements are performed naturally by collective decay. This kind of
''one-pass''teleportation is similar to S. Bose 's proposal \cite{15}, where
the teleportation between two trapped atoms in two independent cavities is
achieved by the leaked cavity photons impinging on a 50-50 beam splitter.
Very recently, Beenakker \emph{et al.} have also theoretically pointed out
that ''one-pass'' teleportation of spin states in quantum Hall system is
possible. \cite{24} The key is the recombination of the electron and hole at
the tunnel barrier. A disadvantage is that the success rate is small.

Just like S. Bose 's protocol, our probabilistic proposal can be modified to
teleportation with insurance, so that in the cases when the protocol is
unsuccessful the original teleported state is not destroyed, but mapped onto
another reserve atom (or dot) $r$. To accomplish this, the key step is the
local redundant encoding of the teleported state \cite{25} before the
collective decay:

\begin{equation*}
\left| \Psi \right\rangle _{code}=\beta (\left| \uparrow \right\rangle
_{2}\left| \downarrow \right\rangle _{r}+\left| \downarrow \right\rangle
_{2}\left| \uparrow \right\rangle _{r})+\alpha (\left| \downarrow
\right\rangle _{2}\left| \downarrow \right\rangle _{r}+\left| \uparrow
\right\rangle _{2}\left| \uparrow \right\rangle _{r}).
\end{equation*}%
If the teleportation is unsuccessful, the coded state is left with either
state $\alpha \left| \uparrow \right\rangle _{r}+\beta \left| \downarrow
\right\rangle _{r}$ or or a state that can be converted $\alpha \left|
\uparrow \right\rangle _{r}+\beta \left| \downarrow \right\rangle _{r}$ by a
known unitary transformation. In this case, one can repeat this procedure
until teleportation is successful.

In summary, we have proposed an alternative way to accomplish quantum
teleportation in solid state charge qubits. Superradiance/subradiance
achieves both entanglement generation and joint measurement during the
teleported process. To the best of our knowledge, it is the first time that
superradiance is pointed out to be useful in quantum teleportation. Our
examples of quantum dots and wires demonstrated possible experimental
realizations, and deserve to be tested with present technologies.

We would like to thank Prof. D. J. Han at NCCU for help discussions. This
work is supported partially by the National Science Council, Taiwan under
the grant numbers NSC 93-2112-M-009-037 and NSC 94-2120-M-009-002.

\end{document}